\documentclass[prl,aps,amssymb,showpacs,twocolumn]{revtex4}
\usepackage{amsmath}
\usepackage{amssymb}
\usepackage{amsthm}
\usepackage{amsfonts}
\usepackage{algorithmic}
\usepackage{enumerate}
\usepackage{latexsym}
\usepackage[dvips]{graphicx}

\newcommand{\beq}{\begin{equation}}
\newcommand{\eneq}{\end{equation}}

\input{epsf}

\begin{document}

\tolerance 10000


\title{Spin Hall Conductance of the Two Dimensional Hole Gas in a Perpendicular Magnetic Field}

\author {Tianxing Ma$^{1}$, Qin Liu$^{1,2}$}

\affiliation{$^{1}$Department of Physics, Fudan University, Shanghai
200433, China\\
 $^{2}$Department of Physics, CCNU, Wuhan 430079, China}
\begin{abstract}
\begin{center}

\parbox{17cm}{The charge and spin Hall conductance of the two-dimensional
hole gas within the Luttinger model with and without inversion
symmetry breaking terms in a perpendicular magnetic field are
studied, and two key phenomena are predicted. The sign of the spin
Hall conductance is modulated periodically by the external magnetic
field, which means a possible application in the future.
Furthermore, a resonant spin Hall conductance in the two-dimensional
hole gas with a certain hole density at a typical magnetic field is
indicated, which implies a likely way to firmly establish the
intrinsic spin Hall effect. The charge Hall conductance is
unaffected by the spin-orbit coupling.}

\end{center}
\end{abstract}

\maketitle
\section{Introduction}
 The intrinsic spin Hall effect (SHE) was predicted by
Murakami {\it et al.}\cite{Sci} in $p$-doped bulk semiconductors of
a Luttinger Hamiltonian and by Sinova {\it et al.}\cite{Sinova} in
two dimensional (2D) electron systems with Rashba spin-orbit
coupling, and the intrinsic SHE has aroused intensive theoretical
studies\cite{Rashba,Zhang,Liu,Smu,B0,SIA,Bernevig,Dai,Loss,Murakami,Shen,Ting,Tao,Spin}
for its potential applications in the emerging field of
spintronics\cite{Apl,Sarmarth,Rmp}. Moreover, experimental
observation of the SHE has also been recently reported in an
electron doped sample with the use of Kerr rotation
microscopy\cite{Kato1,Kato2} and in a two-dimensional hole gas
(2DHG) by angle-resolved polarization detection\cite{Wunderlich}.
Kato {\it et al.}\cite{Kato1} suggested that the
extrinsic\cite{Rashba} SHE in n-type semiconductor was dominant for
that no marked crystal direction dependence was observed in the
strained samples, while the SHE observed in the 2DHG with spin-orbit
coupling was interpreted as intrinsic by Wunderlich {\it et
al.}\cite{Zhang,Liu,Smu,Wunderlich}. Further experimental and
theoretical work is needed to firmly establish the intrinsic VS
extrinsic nature of the SHE. Besides, how to develop spintronics
devices\cite{Apl,Sarmarth,Rmp} is still a very challenging problem.

Although the SHE in the 2DHG has been intensively studied
theoretically, most studies have been limited to zero external
magnetic field\cite{Zhang,Liu,B0,Bernevig,Smu,SIA,Dai,Loss}. The
spin Hall conductance in the 2DHG arises from the splitting between
the light and heavy-hole band and the structural inversion asymmetry
(SIA)\cite{Zhang,SIA} in the 2DHG band. In this paper, we study
transport properties of the 2DHG within the
Luttinger\cite{Zhang,Luttinger1956} model with and without inversion
symmetry breaking terms in a perpendicular magnetic field, and two
key phenomena are reported. First, the sign of the spin Hall
conductance ($i$.$e$, the direction of the spin Hall current) can be
modulated periodically by the external magnetic field, which means a
possible application in the future. Second, a resonant spin Hall
conductance can be observed by adjusting the sample parameters and
magnetic field, and the resonance may be used to firmly establish
the intrinsic SHE in experiments\cite{Dai}. Meanwhile, these two
types of spin-orbit coupling have no effect on the charge Hall
conductance. The rest of the paper is organized as follows: The
model Hamiltonian and the formula for charge and spin Hall
conductance are introduced in section II. Our numerical result and
discussion are shown in section III, and the paper is concluded with
a summary in section IV.
\section{Theoretical framework}
The effective Hamiltonian we consider for a 2DHG is a sum of both
Luttinger and spin-$\vec{S} = 3/2$ SIA
terms\cite{Zhang,Luttinger1956}:
\begin{eqnarray}
H_0 &=& \frac{1}{2m} (\gamma_1 + \frac{5}{2} \gamma_2) p^2 -
\frac{\gamma_2}{m} (p_x^2S_x^2+ p_y^2S_y^2 +{\langle p_z^2 \rangle}
S_z^2) \nonumber \\
&-& 2\frac{\gamma_2}{m}\{p_x,p_y\}\{S_x,S_y\}+\alpha (\vec{S} \times
\vec{p}) \cdot{\hat{z}}
\end{eqnarray} where $\vec{S}$ is spin-$\frac{3}{2}$
operator and $m$ is the bare electron mass, and we define $\{A,B\}
=\frac{1}{2}(AB+BA)$, $p^2=p_x^2+ p_y^2+{\langle p_z^2 \rangle}$. In
addition, $\gamma_{1}$ and $\gamma_{2}$ are two dimensionless
parameters modeling the effective mass and spin-orbit coupling
around the $\Gamma$ point. The confinement of the well in the $z$
direction quantizes the momentum on this axis and it is approximated
by the relation $\langle p_z \rangle=0, {\langle p_z^2 \rangle}
\approx (\pi \hbar/ d)^2$ for a quantum well with thickness $d$.
This is not the most general Hamiltonian when a magnetic field is
present. The most general one involves both linear and cubic spin
Zeeman terms, and we neglect these terms in Hamiltonian (1), which
is a good approximation for GaAs\cite{Winkler,Ma}.

We impose a magnetic field $\vec{B}=B\hat{z}$ by choosing the Landau
gauge $\vec{A}=-yB\hat{x}$, and then $p_{x}= \hbar
k_{x}+\frac{eB}{c}y$, where $-e$ is the electric charge. The
destruction operator\cite{Luttinger1956}
$a=\frac{1}{\sqrt{2m\hbar\omega}}(p_{x}+ip_{y})$, and the creation
operator $a^{\dag }=\frac{1}{\sqrt{2m\hbar\omega}}(p_{x}-ip_{y})$
are introduced to describe Landau levels, where
$\omega=\frac{eB}{mc}$. These operators have the commutation
$[a,a^{\dag }]=1$. In terms of these operators and using explicit
matrix notation with $S=\frac{3}{2}$ eigenstates in the order
$S_{z}=+\frac{3}{2}$, $+\frac{1}{2}$, $-\frac{1}{2}$,
$-\frac{3}{2}$, Hamiltonian (1) within a magnetic field can be
rewritten as
\begin{equation}
H_{0}=\hbar \omega \left(
\begin{array}{cccc}
H_{11} & i\sqrt{3}\eta a^{\dag } & -\sqrt{3}\gamma _{2}a^{\dag2} & 0 \\
-i\sqrt{3}\eta a & H_{22} & 2i\eta a^{\dag } & -\sqrt{3}\gamma _{2}a^{\dag 2} \\
-\sqrt{3}\gamma _{2}a^{2} & -2i\eta a & H_{22} & i\sqrt{3}\eta a^{\dag } \\
0 & -\sqrt{3}\gamma _{2}a^{2} & -i\sqrt{3}\eta a & H_{11}
\end{array}
\right), \nonumber \\
\end{equation}
with
\begin{eqnarray}
H_{11}&=&(\gamma_{1}+\gamma_{2})(a^{\dag }a+\frac{1}{2})+
\frac{\beta}{2}(\gamma_1 -2\gamma_2), \nonumber \\
H_{22}&=&(\gamma_{1}-\gamma_{2})(a^{\dag }a+\frac{1}{2})+
\frac{\beta}{2}(\gamma_1 +2\gamma_2).
\end{eqnarray}
where the dimensionless parameters $\eta=\alpha
m\sqrt{\frac{c}{2\hbar eB}}$, and
$\beta=\frac{\pi^{2}\hbar}{d^{2}m\omega}$.

Depending on the confinement scale $d$ the Luttinger term is
dominant for $d$ not too small, while the SIA term becomes dominant
for infinitely thin wells, which corresponds to high junction
fields. We first consider the case when the confinement scale $d$ is
not too small, that is, neglect the SIA term:
\begin{eqnarray}
H^{L}_0&=& \frac{1}{2m} (\gamma_1 + \frac{5}{2} \gamma_2) p^2 -
\frac{\gamma_2}{m} (p_x^2S_x^2+ p_y^2S_y^2 +{\langle p_z^2 \rangle}
S_z^2)\nonumber \\
&-& 2\frac{\gamma_2}{m}\{p_x,p_y\}\{S_x,S_y\}.
\end{eqnarray}
The eigenstate of $H^{L}_{0}$ can be taken as
\begin{equation}
\left\vert n,s,f\right\rangle =\left(
\begin{array}{c}
C_{nsf1}\phi _{n} \\
C_{nsf2}\phi _{n-1} \\
C_{nsf3}\phi _{n-2} \\
C_{nsf4}\phi _{n-3}
\end{array}%
\right),
\end{equation}%
where $\phi _{n}$ is the eigenstate of the n$th$ Landul level in the
absence of the Luttinger interaction, and the eigenvalues in units
of $\hbar\omega$ when $n\geq3$ can be obtained analytically as
\begin{widetext}
\begin{eqnarray}
E_{n+1sf}=(n+\frac{1}{2} \beta )\gamma _{1}+\frac{f}{2}(\gamma
_{1}+2\gamma _{2})+s\sqrt{\gamma _{1}^{2}+[2f(n-\beta)+1]\gamma
_{1}\gamma
_{2}+[(4n-2\beta)(n+f)+f\beta+\beta^{2}+\frac{1}{4}]\gamma
_{2}^{2}},\\
\theta _{n+1sf}=\arctan
(\frac{-\sqrt{3}\gamma_{2}\sqrt{n(n+f)}}{[\gamma _{1}+f(n-\beta
)\gamma
_{2}\allowbreak +\frac{\gamma _{2}\allowbreak }{2}\allowbreak ]+s\sqrt{%
\gamma _{1}^{2}+[2f(n-\beta )+1]\gamma _{1}\gamma _{2}+[(4n-2\beta
)(n+f)+\beta ^{2}+f\beta +\frac{1}{4}]\gamma _{2}^{2}}}).
\end{eqnarray}
\end{widetext}
The eigenvectors for $s=\pm1$, $f=1$ can be expressed as
\begin{eqnarray}
\left(
\begin{array}{c}
\cos \theta _{nsf} \\
0 \\
\sin \theta _{nsf} \\
0%
\end{array}%
\right), \end{eqnarray} while those for $s=\pm1$, $f=-1$ can be
expressed as
\begin{eqnarray}
\left(
\begin{array}{c}
0 \\
\cos \theta _{nsf} \\
0 \\
\sin \theta _{nsf}%
\end{array}%
\right), \end{eqnarray} where $\theta _{nsf}$ is defined in Eq.(6),
and $\theta_{n,1,1}=\theta_{n,-1,1}+\frac{\pi}{2}$,
$\theta_{n,1,-1}=\theta_{n,-1,-1}-\frac{\pi}{2}$. The magnitude of
$\theta_{nsf}$ is important and we will discuss it corresponding to
the parameters used in this paper bellow, and we can simply define
that $\theta_{n,-1,\pm1}\in(0,\frac{\pi}{2})$ here. From large n
limit, we can deduce that states $\left\vert n,+1,\pm1\right\rangle$
indicate light-hole bands (LH$^{\pm}$) and $\left\vert
n,-1,\pm1\right\rangle$ indicates heavy-hole bands
(HH$^{\pm}$)\cite{Zhang,Ma}. Besides, the eigenstates and
eigenvalues for n$<3$ are easy to obtain and we do not show them
explicitly here.
\begin{figure}
\includegraphics[scale=0.5]{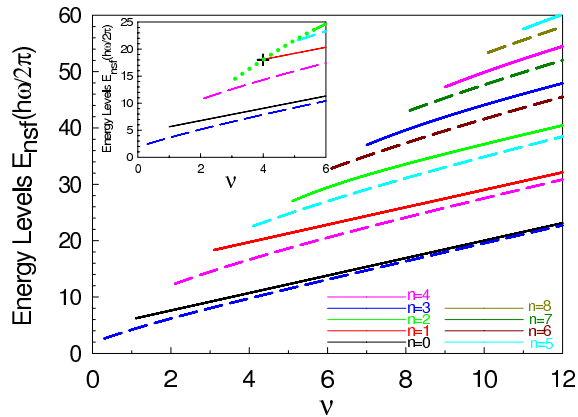}
\caption{(Color online) Landau levels (units: $\hbar\omega$) as
functions of filling factors $\nu$. Parameters used are taken from
Ref.\cite{Zhang,Wunderlich} and $\alpha=0$. Different colors denote
different n and only energy levels occupied for corresponding $\nu$
are shown. Solid lines indicate Landau levels of mostly holes with
spin-$\frac{3}{2}$ and dashed lines indicate Landau levels of mostly
holes with spin-$-\frac{3}{2}$. Inset: $n_{h}=2.73\times
10^{16}/m^{2}$. The dotted line indicates the Landau level of mostly
holes with spin-$-\frac{1}{2}$. The cross indicates the energy
crossing which gives rise to a resonance in spin Hall conductance.}
\end{figure}
The energy levels as functions of the filling factor $\nu
=\frac{N_{h}}{N_{\phi}}=\frac{n_{h}2\pi\hbar c}{eB}$ are shown in
Fig.1. We use different color lines to denote different n and in
order to give a more clear illumination bellow, we only plot the
energy levels occupied for every $\nu$. Solid lines indicate Landau
levels of mostly holes with spin-$\frac{3}{2}$ (states $\left\vert
n,-1,1\right\rangle $) and dashed lines indicate Landau levels of
mostly holes with spin-$-\frac{3}{2}$ (states $\left\vert
n,-1,-1\right\rangle $). From the bottom up, n increases one by one
for solid lines (3$\sim$8) and dashed lines (0$\sim$5) respectively.
Parameters used are taken from Ref.\cite{Wunderlich},
$n_{h}=2.0\times 10^{16}/m^{2}$, $\gamma_{1}=6.92, \gamma_{2}=2.1$,
$2\frac{\gamma_{2}}{m}(\frac{\pi\hbar}{d})^{2}$=40meV, and
$d$=8.3nm. In addition, $\alpha=0$. Inset: all the parameters used
are the same except $n_{h}=2.73\times 10^{16}/m^{2}$. The dotted
line indicates Landau levels of mostly holes with
spin-$-\frac{1}{2}$. The cross indicates the energy crossing which
we will discuss below.

Let's begin to study the law in Fig.1 in detail (Don't include the
Fig inset). When $\nu_{i}=$ odd integers, the top energy levels
occupied can be written as
\begin{eqnarray}
E^{\nu_{i}=odd}_{occupied}=E_{n=\frac{\nu_{i}}{2}+\frac{5}{2},-1,-1},
\end{eqnarray}
while, when $\nu_{i} =$ even integers, the top energy levels
occupied are
\begin{eqnarray}
E^{\nu_{i}=even}_{occupied}=E_{n=\frac{\nu_{i}}{2}-1,-1,1},
\end{eqnarray}
and the total energy levels occupied for every $\nu$ include both
states when $\nu_{i}=$odd and $\nu_{i}=$even where
$\nu_{i}=1,2,...\nu$. We should add that when $n<3$, the definition
of $\left\vert n,s,f\right\rangle $ is not exact, so we simply take
$\left\vert 2,-1,1\right\rangle$ as the lowest energy level of $n=2$
and $\left\vert 1,1,-1\right\rangle$ indicates the lowest energy
level of $n=1$ in the whole paper. It is interesting to find that
these energy levels arrange regularly, and the Landau levels of
mostly holes with spin-$\frac{3}{2}$ and mostly holes with
spin-$-\frac{3}{2}$ appear by turns. This law is important and it
will determine the behavior of the spin Hall conductance which we
will discuss below.

Now we turn to study the system in the presence of an electric field
applied in $-\vec{y}$ direction. The full Hamiltonian $H$ is
\begin{eqnarray}
H= H^{L}_{0}+H',
\end{eqnarray}
with
\begin{eqnarray}
H^{'}= eEy=eE\sqrt{\frac{\hbar c}{2eB}}(a^{\dag}+a)-\frac{E\hbar
c}{B}k_{x}.
\end{eqnarray}
By treating $H^{'}$ as a perturbation term\cite{Shen} up to the
first order, we calculate the charge and spin Hall conductance along
$x$ direction, which is the most interesting case.

The charge current operator for a hole along $x$ direction is given
by $j_{c}=e\upsilon_{x}$ with $\upsilon_{x}=\frac{1}{i\hbar}[x,H]$,
and the spin $S^{z}$ component current operator can be expressed
as\cite{Murakami}
\begin{eqnarray}
j_{s}&=\frac{\hbar}{3}\{S^{z},\upsilon_{x}\}.
\end{eqnarray}
From Eq.(13) and eigenstates (4) used, we can find that the
contribution to the charge and spin Hall conductance arising from
the zeroth order in $H$ is zero for  $<n,s,f|j_{c,s}|n,s,f>=0$.

We shall only focus on the first order in the perturbation in
$H^{'}$, which is
\begin{eqnarray}
(j_{c,s})_{nsf}&=&\sum_{n^{\prime }s^{\prime }f^{\prime }}\frac{
\left\langle n,s,f\right\vert j_{c,s}\left\vert n^{\prime
},s^{\prime },f^{\prime }\right\rangle }{\epsilon
_{nsf}-\epsilon _{n^{\prime }s^{\prime }f^{\prime }}} \nonumber \\
&\times&\left\langle n^{\prime },s^{\prime },f^{\prime }\right\vert
H^{\prime }\left\vert n,s,f\right\rangle+H.c.,
\end{eqnarray}
where $(j_{c,s})_{nsf}$ is the current carried by a hole in the
state $|n,s,f>$ of $H^{L}_{0}$, and $\epsilon _{nsf}=\hbar\omega E
_{nsf}$. In addition, $n'=n\pm 1$. The average current density of
the system is given by
\begin{eqnarray}
I_{c,s}&=&\frac{1}{L_{y}}\sum_{nsf}(j_{c,s})_{nsf}f(\epsilon
_{nsf}),
\end{eqnarray}
where $f(\epsilon _{nsf})$ is the Fermi distribution function, and
the charge or spin hall conductance is
\begin{eqnarray}
G_{c,s}&=&\frac{I_{c,s}}{EL_{x}}.
\end{eqnarray}
We can finally obtain the complete charge and spin Hall conductance,
and the 2DHG system we consider is confined in the $x-y$ plane of an
area $L_{x}\times L_{y}$.

Our calculation of the charge and spin Hall conductance is complete
for the linear response theory because that the zeroth order mean
value of $j_{c,s}$ is zero in each state $|n,s,f>$, meanwhile we
estimate the effect of relaxation to the Fermi distribution function
$f(\epsilon _{nsf})$ by means of Boltzman equation. The extra term
of $f(\epsilon _{nsf})$ from relaxation is proportional to the
applied field, thus the effect on the conductance from relaxation is
second order of the applied field $E$, so we can roughly neglect it
at linear response region.
\section{Results and discussions}
\begin{figure}
\includegraphics[scale=0.6]{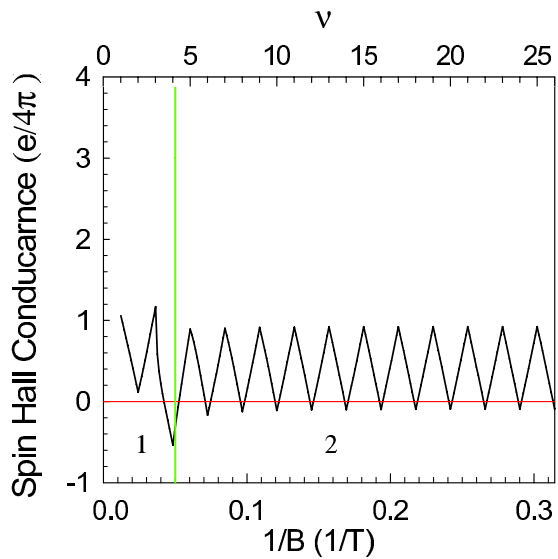}
\caption{(Color online) Spin hall conductance versus $1/B$ at $T=0$,
and all the parameters used are the same as those in Fig.1. }
\end{figure}

We calculate the charge Hall conductance at $T=0$ numerically
firstly. Comparing with the usual quantum Hall effect case, $i$.$e,
G_{c}=\nu\frac{e^{2}}{h}$, our results show that the Luttinger type
spin-orbit coupling has no effect on the charge Hall conductance,
and this result is consistent with the quantization of the Hall
conductance\cite{Hall}.

Now let's turn to discuss the magnetic field dependence of the spin
Hall conductance. We calculate the spin Hall conductance within
Hamiltonian (11), and the spin Hall conductance as a function of
$1/B$ is shown in Fig.2. In order to be convenient to illustrate it,
Fig.2 is divided into two regions by a vertical green line at the
value of magnetic field $B=20T$. At the region that $B<20T$ (region
2), the sign of the spin Hall conductance changes periodically with
the decreasing of magnetic field, and the periodicity is
$\frac{e}{n_{h}hc}$. The spin Hall conductance oscillates as a
result of the alternative occupation of mostly holes with
spin-$\frac{3}{2}$ and mostly holes with spin-$-\frac{3}{2}$. It
reaches its maxima where $\nu$= odd and minima where $\nu$=even,
while the sign of the spin Hall conductance is opposite for odd and
even $\nu$, except when $\nu=2$.

Let's devote our attention to the region 1, especially where $\nu$
changes from $3$ to $5$, and the corresponding $B$ is $16.5\sim
27.6T$, which can be achieved in laboratory now. When the absolute
value of the spin Hall conductance is larger than 0.3 ($i$.$e$, the
spin Hall current is large enough to be detected) in the $-x$
direction, the range of magnetic field is $20.0\sim 21.5T$, while in
the $+x$ direction, the value of magnetic field is in the range of
$15.2\sim 18.3T$. The range of magnetic field for a notable spin
Hall current is wide enough, which makes it easy to modulate the
direction of the spin Hall current. If the direction of the spin
Hall current can be detected, we can take the direction of the spin
Hall current as a new sign, i.e, $+x$ direction means ``0" and $-x$
direction means ``1", and these two states maybe the basic for a new
logic electronic-device. In addition, we have made an scan on hole
density, as well as magnetic field dependence of sign changes. We
find that such rich sign changes are robust even in moderate
magnetic fields, however, a more notable sign changes is sensitive
to the hole density\cite{Ma}. The sign issue should be regarded as a
very important aspect of SHE and can be used in future
experiments\cite{Ma,Yao}.

Bernvig $et$ $al.$\cite{Zhang} have calculated the spin Hall
conductance within the same Hamiltonian (3) in the absence of a
magnetic field, and their result shows that the spin hall
conductance is $\frac{1.9e}{8 \pi}$, which is in good agreement with
the numerical estimate by Wunderlich $et$ $al.$\cite{Wunderlich}. As
shown in Fig.2, the jump in the spin Hall conductance tends to a
stable value of $\frac{2.0e}{8 \pi}$, which is very near to the
universal value.
\begin{figure}
\includegraphics[scale=0.6]{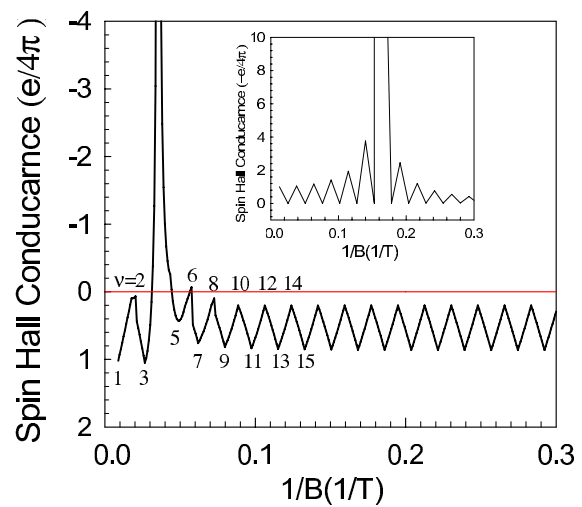}
\caption{(Color online) Resonant spin Hall conductance versus $1/B$
at $T=0$, and all the parameters used are the same as those in Fig.2
except $n_{h}=2.73\times 10^{16}/m^{2}$. Inset: Spin Hall
conductance in 2DEG with a Rashba spin-orbit coupling versus $1/B$
at T=0, and data are achieved within the method in Ref\cite{Shen}.}
\end{figure}

In the 2DEG with a Rashba spin-orbit coupling in a perpendicular
magnetic field, the resonant spin Hall conductance stems from energy
crossing of different Landau levels near the Fermi level\cite{Shen}.
Although this kind of energy crossing can appear when
$n_{h}=2.0\times 10^{16}/m^{2}$ of the sample we consider, but
energy levels $\epsilon_{n=1,1,-1}$ and $\epsilon_{n'=n+1,-1,1}$ are
higher than Fermi level as shown in Fig.1, so they do not contribute
to the spin Hall conductance, and then there's no resonant spin Hall
conductance when $n_{h}=2.0\times 10^{16}/m^{2}$ as shown in Fig.2.
If we modulate the hole density $n_{h}$, and take a value when the
energy crossing occurs near the Fermi level as shown in Fig.1 inset
($n_{h}=2.73\times 10^{16}/m^{2}$), there shall be an effective
energy crossing between $\epsilon_{n=1,1,-1}$ and
$\epsilon_{n'=n+1,-1,1}$ which occurs at $B_{r}=28.52T$ (the key
point denoted by a cross in Fig.1 inset), we shall see a resonance
in the spin Hall conductance as shown in Fig.3.

In order to make the energy crossing to be occurred near the Fermi
level, it should request
\begin{eqnarray}
3<\frac{n_{h}2\pi\hbar}{eB_{r}}<4,
\end{eqnarray}
that is, when $2.07\times 10^{16}/m^{2}< n_{h}<2.75\times
10^{16}/m^{2}$, the ``effective" energy crossing between mostly
holes with spin-$-\frac{1}{2}$ and holes with spin-$\frac{3}{2}$
shall give rise to a resonant spin Hall conductance in $-x$
direction as shown in Fig.3. The resonant spin Hall conductance
means spin accumulation\cite{Nomura,Tao,Spin} might be observed in
experiments near the edges of a semiconductor channel, and it may be
a way to distinguish the intrinsic from the extrinsic SHE\cite{Dai}.

Even all the results above are obtained numerically, we can discuss
them in some analytical way. The states near Fermi surface are
mostly $|n,-1,\pm1>$. Let's argue the contribution for the spin Hall
conductance when $\nu_{i}$=odd firstly, and the key states are
$|n,-1,-1>$. We can learn from the properties of eigenvectors shown
in Eq.(7) and (8) that only states $|n\pm1,\pm1,-1>$ shall
contribute to the spin Hall conductance for $\left\langle n^{\prime
},\pm1,+1\right\vert H^{\prime }\left\vert n,-1,-1\right\rangle=0$,
furthermore, $|\epsilon_{n,-1,-1}-\epsilon_{n\pm1},-1,-1|\ll
|\epsilon_{n,-1,-1}-\epsilon_{n\pm1,1,-1}|$, so the spin Hall
conductance arising from states $|n,-1,-1>$ shall be dominated by
states $|n\pm1,-1,-1>$
\begin{eqnarray}
(j_{S}^{odd})_{n,-1,-1}&=&\sum_{n^{\prime }s^{\prime }f^{\prime
}}\frac{ \left\langle n,-1,-1\right\vert j_{s}\left\vert n^{\prime
},s^{\prime },f^{\prime }\right\rangle }{\epsilon
_{n,-1,-1}-\epsilon
_{n^{\prime }s^{\prime }f^{\prime }}}\nonumber \\
 &\times&\left\langle n^{\prime
},s^{\prime },f^{\prime }\right\vert H^{\prime }\left\vert
n,-1,-1\right\rangle
+H.c.\nonumber \\
&\approx&\frac{Ec\hbar}{6B}\sum_{n^{\prime
},-1,-1}g_{S}^{odd}(n\rightarrow n^{\prime }),
\end{eqnarray}
and then all the part for $\nu_{i}$=odd can be obtained as
\begin{eqnarray}
G_{\nu_{i}=odd}\approx\frac{e}{12\pi}g_{S}^{odd}(\frac{\nu_{i}}{2}+\frac{5}{2}+1)f(\epsilon
_{\frac{\nu_{i}}{2}+\frac{7}{2},-1,-1}),
\end{eqnarray}
where the Landau degeneracy factor $m\omega/(2\pi\hbar)$ is also
included and
\begin{eqnarray}
g_{S}^{odd}(n\rightarrow n')&=&\frac{6B}{Ec\hbar}\frac{\left\langle
n,-1,-1\right\vert j_{s}\left\vert n',-1,-1\right\rangle}{\epsilon
_{n,-1,-1}-\epsilon_{n',-1,-1}}\nonumber \\
 &\times& \left\langle
n',-1,-1\right\vert H^{\prime }\left\vert n,-1,-1\right\rangle+H.c\nonumber \\
g_{S}^{even}(n\rightarrow n')&=&\frac{6B}{Ec\hbar}\frac{\left\langle
n,-1,1\right\vert j_{s}\left\vert n',-1,1\right\rangle}{\epsilon
_{n,-1,1}-\epsilon_{n',-1,1}} \nonumber \\
&\times& \left\langle n',-1,1\right\vert H^{\prime }\left\vert
n,-1,1\right\rangle+H.c.
\end{eqnarray}
In deriving Eq.(18), we have used the fact that the transitions
$|i,-1,-1>\rightarrow |i-1,-1,-1>$ is canceled by
$|i-1,-1,-1>\rightarrow |i,-1,-1>$ where $i\leq n$, so only the
$g_{S}^{odd}(n\rightarrow n+1)$ is reserved and we simply write it
as $g_{S}^{odd}(n+1)$. From Eq.(9), we can obtain that
$n+1=\frac{\nu}{2}+\frac{5}{2}+1$. In the similar way, all the parts
for $\nu_{i}$=even can be written as
\begin{eqnarray}
G_{\nu_{i}=even}&\approx&[\frac{e}{12\pi}g_{S}^{even}(\frac{\nu_{i}}{2})f(\epsilon
_{\frac{\nu_{i}}{2},-1,+1})]_{\nu_{i}\geq 4},
\end{eqnarray} and $G_{\nu_{i}=2}=-\frac{e}{4\pi}$ can be calculated
immediately. At zero temperature, when occupied, any
$f(\epsilon_{occupied})=1$. The whole spin Hall conductance can be
expressed as
\begin{eqnarray}
G_{\nu=odd}&\approx&\frac{e}{12\pi}[g_{S}^{even}(\frac{\nu-1}{2})+g_{S}^{odd}(\frac{\nu}{2}+\frac{7}{2})],\nonumber \\
G_{\nu=even}&\approx&\frac{e}{12\pi}[g_{S}^{even}(\frac{\nu}{2})+g_{S}^{odd}(\frac{\nu}{2}+3)].
\end{eqnarray}

The magnitude of $\theta_{n,-1,\pm1}$ is important. Corresponding to
the parameters used, and for the energy levels occupied of every
$\nu$, we can define that $\theta_{n,-1,1}\in(0,\frac{\pi}{4})$ and
$\theta_{n,-1,-1}\in(\frac{\pi}{4},\frac{\pi}{2})$. From the
numerical calculation, we can obtain the following linear dependence
of filling factor that when $\nu=odd$,
\begin{eqnarray}
g_{S}^{odd}(n)&\approx&1.12n-2.07, g_{S}^{even}(m)\approx-1.12m-0.97, \nonumber \\
n&=&\nu/2+5/2, m=(\nu-1)/2-1
\end{eqnarray}
and when $\nu=even$,
\begin{eqnarray}
g_{S}^{odd}(n)&\approx&1.12n-2.04, g_{S}^{even}(m)\approx-1.12n-0.88, \nonumber \\
n&=&(\nu-1)/2+5/2, m=\nu/2-1.
\end{eqnarray}
From Eqs.(7) and (8), we can also learn that $g_{S}^{odd}$ arises
from the transition between mostly holes with spin-$-\frac{3}{2}$
and $g_{S}^{even}$ arises from the transitions between mostly holes
with spin-$\frac{3}{2}$. The oscillations on the spin Hall
conductance due to the alternative occupation of mostly holes with
spin-$\frac{3}{2}$ and mostly holes with spin-$-\frac{3}{2}$ can be
explained approximately, and our analytical process also indicates
that the contribution from the transitions between mostly holes with
spin-$\frac{3}{2}$ (or spin-$-\frac{3}{2}$) should be the dominant
part of the spin Hall conductance.

The contribution to the resonant spin Hall conductance can be
obtained as
\begin{eqnarray}
g_{r}&=&[3\sqrt{2}(\gamma _{1}+\gamma _{2})\cos \theta
_{2,-1,1}-2\sqrt{3}\gamma_{2} \sin \theta _{2,-1,1}]\nonumber \\
 &\times&
\frac{\sqrt{2}\cos \theta _{2,-1,1}}{E_{2,-1,1}-E_{1,1,-1}},
\end{eqnarray}
where $\theta
_{2,-1,1}=\arctan\frac{-\sqrt{6}\gamma_{2}}{[\gamma_{1}+(\frac{3}{2}-\beta)\gamma_{2}]-\sqrt{[\gamma
_{1}+(\frac{3}{2}-\beta)\gamma_{2}]^{2} +6\gamma _{2}^{2}}}$, and
$E_{2,-1,1}-E_{1,1,-1}<0$ which leads to a resonant spin Hall
conductance in $-x$ direction. A more detailed discussion will be
given in our further work\cite{Ma}.

\begin{figure}
\includegraphics[scale=0.6]{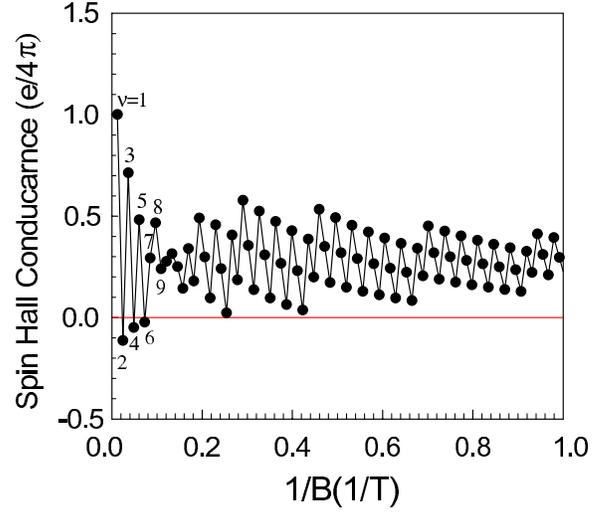}
\caption{(Color online) Spin Hall conductance versus $1/B$ at $T=0$,
Parameters used are the same as those in Fig.1 except
$\alpha=1.0\times 10^{5}m/s$.}
\end{figure}
The relatively large 5 meV measured spitting\cite{Wunderlich,Zhang}
of the HH band implies that the effect of SIA terms is important. We
have computed the spin conductance within the Hamiltonian (2) in the
presence of an electric field in $-\vec{y}$ direction, and the
result is shown in Fig.4.

When $\nu>7$($i$.$e$, $B<11.8T$), the spin Hall conductance
oscillates as a result of the alternative occupation of mostly holes
with spin-$-\frac{3}{2}$, spin-$\frac{3}{2}$ and spin-$-\frac{3}{2}$
between some turning points, $\nu=$7, 10, 37, 56, 76, 96, while the
spin Hall conductance oscillates due to the alternative occupation
of mostly holes with spin-$\frac{3}{2}$ and spin-$-\frac{3}{2}$ when
$B>11.8T$. Furthermore, the turning points appear periodically when
$B<8.3T$, and the period approaches a stable value
19$\frac{e}{n_{h}h c}$ when $B<2.2T$. The difference between Fig.2
and Fig.4 can be found from the effect of SIA term. The SIA term
arising from the structure inversion symmetry rearrange the energy
levels when $\nu>7$\cite{Ma}, and this effect is similar to the case
in the absence of magnetic field\cite{Zhang,Dai}.

The charge Hall conductance has also been studied within the
Hamiltonian (2) in the presence of an electric field in $-\vec{y}$
direction, and our result shows that the SIA term has no effect on
the charge Hall conductance. Our further calculation shows that we
can also predict a resonant spin Hall conductance within the
Luttinger model including the SIA term at a typical magnetic
field\cite{Ma} by adjusting the hole density.

After the completion of this paper, we become aware of the
independent work\cite{Zarea} of M. Zarea and S. E. Ulloa, which
studies the system of 2D heavy holes described by a $k$-cubic model
in a magnetic field. In the thin quantum well limit, LH bands become
energetically irrelevant and the HH bands can be effectively
described by the 2D HH system\cite{Liu,SIA,Loss,Sch} where only the
lowest HH subband is occupied. This condition can be satisfied when
the quasi 2D system is sufficiently narrow and the density and the
temperature are not too high\cite{Liu,Loss,Sch}. Although both the
paper by M. Zarea {\it et al.} and this paper by us compute the spin
hall conductance in 2DHG in a magnetic field, M. Zarea {\it et
al.}\cite{Zarea} focuses on the low field regime\cite{Sch}. In
strong magnetic field regime and at a high hole density, the
resonance reported in this paper can not be achieved within the two
dimensional HH system\cite{Ma}, which is due to the interplay
between mostly holes with spin-$-\frac{1}{2}$ and holes with
spin-$\frac{3}{2}$. Moreover, the resonance remains within Luttinger
model even including Zeeman splitting, and the resonance effect
stemming from energy crossing of different Landau levels near the
Fermi level due to the competition of Zeeman energy splitting and
$k$-cubic spin-orbit coupling is out of reach for either real
materials or in experiments\cite{Ma}. However, when only the lowest
HH subband is occupied, the 2D HH system can reproduce qualitatively
the result within Luttinger model with Rashba spin-orbit coupling in
the presence of a magnetic field\cite{Ma}.

\section{Summary}
In summary, we have studied the charge and spin Hall conductance in
p-type GaAs quantum well structure described by a Luttinger
Hamiltonian with the SIA term in a perpendicular magnetic field. The
sign of the spin Hall conductance ($i$.$e$, the direction of the
spin Hall current) purely caused by the Luttinger type spin-orbit
coupling can be modulated periodically by the external magnetic
field, which means a possible application in the future. The
effective energy crossing between mostly holes with
spin-$-\frac{1}{2}$ and holes with spin-$\frac{3}{2}$ at a typical
magnetic field gives rise to a resonant spin Hall conductance with a
certain hole density, and the jump in the spin Hall conductance is
very near to the universal value of $\frac{1.9e}{8 \pi}$. Our
calculation shows that the hole density, the well thickness and the
value of the magnetic field for the resonance are accessible in
experiments. The resonance may be used to distinguish the intrinsic
SHE from the extrinsic one. Although the SIA term rearranges the
energy level of the Luttinger hamiltonian, the spin Hall conductance
is similar to the case caused by pure Luttinger type spin-orbit
coupling. The dominate contribution to the spin Hall conductance is
the transitions arising from $|n,-1,-1>$ to $|n\pm1,-1,-1>$ (between
mostly holes with spin-$-\frac{3}{2}$) and $|n,1,-1>$ to
$|n\pm1,1,-1>$ (between mostly holes with spin-$\frac{3}{2}$). Our
results also show that both Luttinger and spin-$\vec{S} = 3/2$ SIA
spin-orbit coupling have no effect on the charge Hall conductance.

\acknowledgments We thank Professors Shou-Cheng Zhang, Ruibao Tao,
Shun-Qing Shen, Wuming Liu, Yu Shi and Dr. B. Andrei Bernevig for
many helpful discussions.

\end{document}